\documentclass[aps,prb,twocolumn,superscriptaddress,floatfix]{revtex4-2}
\usepackage{amsmath,amsfonts,amssymb,amsthm}
\usepackage[T1]{fontenc}
\usepackage[utf8]{inputenc}
\usepackage{bbm,bm,bbold}
\usepackage{booktabs}
\usepackage{multirow}
\usepackage{hhline}
\usepackage{comment}
\usepackage{float}
\usepackage{color}
\usepackage{graphicx,latexsym}
\usepackage[colorlinks=true,urlcolor=blue,citecolor=blue,linkcolor=blue]{hyperref}

\usepackage{nameref}
\usepackage{varioref}
\usepackage{cleveref}

\begin{document}

\renewcommand{\vec}{\mathbf}
\renewcommand{\Re}{\mathop{\mathrm{Re}}\nolimits}
\renewcommand{\Im}{\mathop{\mathrm{Im}}\nolimits}
\newcommand\scalemath[2]{\scalebox{#1}{\mbox{\ensuremath{\displaystyle #2}}}}

\title{Fast Hubbard-Thouless pumping with local counterdiabatic driving}

\author{Alexander Chansky}
\affiliation{Universit\'{e} de Strasbourg, Institut de Physique et Chimie des Mat\'{e}riaux de Strasbourg, Strasbourg, France}
\affiliation{Department of Physics, Boston University, 590 Commonwealth Ave., Boston, MA 02215, USA}
\email{alexander.zalischansky@etu.unistra.fr}

\begin{abstract}
Counterdiabatic (CD) driving suppresses the excitations of fast protocols by adding an
auxiliary Hamiltonian, but in many-body and topological systems the exact CD term is
generally non-local and multi-body. We study this obstruction for Thouless pumping in the
Rice-Mele model, first without and then with interactions. For noninteracting fermions,
the exact CD term of the pump is non-local, but the variational framework of Sels and
Polkovnikov yields accurate nearest-neighbor approximations. For the spinful Rice-Mele
chain with onsite and nearest-neighbor repulsion, in the regime where the competition of
the two interactions stabilizes quantized pumping, this approach fails: the
infinite-temperature variational action only penalizes the CD amplitudes, and the
resulting two-parameter nearest-neighbor term does not prevent the pump from breaking down
when the cycle time is reduced from $50/t$ to $2/t$. We trace the failure to the structure
of the exact many-body gauge potential, whose amplitude for each hop depends on the
interaction energy that the hop costs. Resolving the nearest-neighbor currents by this
energy, and fixing the coefficients with a ground-state-weighted action, gives a local
ansatz with 14 coefficients that keeps the interacting pump quantized for cycle times
$T=2/t$ and $T=1/t$, with an accuracy per site that does not degrade between rings of 8
and 12 sites.
\end{abstract}

\date{\today}
\maketitle
\section{Introduction}
\label{sec:intro}

The precise control of quantum states is a foundational requirement for advancements in quantum computing and the exploration of topological phases of matter. When a system is driven by a time-dependent Hamiltonian, the adiabatic theorem guarantees that it will remain in its instantaneous ground state provided the evolution is sufficiently slow. Strictly adiabatic protocols are, however, often experimentally impractical: lengthy evolution times leave the system vulnerable to decoherence, environmental noise, and uncontrolled dissipation~\cite{bukovUniversalHighFrequencyBehavior2015}.

Counterdiabatic (CD) driving, or transitionless quantum driving~\cite{demirplakAdiabaticPopulationTransfer2003,berryTransitionlessQuantumDriving2009,gueryodelinShortcutsAdiabaticityConcepts2019}, presents a powerful framework to overcome these limitations. By augmenting the original Hamiltonian with an auxiliary gauge potential, CD driving exactly suppresses the non-adiabatic transitions that occur during rapid evolution. While this approach is elegant and readily applied to simple two-level systems, extending it to complex many-body and topological systems introduces profound physical constraints: constructing exact CD operators for systems undergoing continuous quantum phase transitions often demands highly non-local, multi-body interactions that are impossible to realize in a laboratory setting.

Bridging this gap between exact CD driving and experimental accessibility has been pursued along several complementary lines. For few-level systems, the non-local CD term can be synthesized via a Floquet drive built from local controls~\cite{boyersFloquetengineeredQuantumState2019}; this strategy extends to many-body systems through a systematic nested-commutator expansion of the gauge potential~\cite{claeysFloquetEngineeringCounterdiabaticProtocols2019}. For continuous quantum phase transitions, del Campo \emph{et al.}~\cite{delcampoAssistedFiniteRateAdiabatic2012} showed that the CD coupling range in the quantum Ising chain diverges at criticality, a result we use below as a consistency check for the equivalent Kitaev chain. Closest to the present work, two recent studies apply CD driving to the Rice-Mele model to accelerate Thouless pumping: Ref.~\cite{chielDynamicallyRobustCounterdiabatic2026} obtains an exact, local CD Hamiltonian only in a fine-tuned ``control-freak'' limit of the model (supplemented by numerical optimization away from it), while Ref.~\cite{liuShortcutsAdiabaticTopological2025} uses the exact, generally non-local CD term directly, realized through internal atomic levels rather than real-space locality. The present work instead asks whether the systematic, variational gauge-potential expansion of Sels and Polkovnikov \cite{selsMinimizingIrreversibleLosses2017} -- already used to construct local CD approximations in a range of other many-body settings~\cite{kolodrubetzGeometryNonadiabaticResponse2017,claeysFloquetEngineeringCounterdiabaticProtocols2019,xieVariationalCounterdiabaticDriving2022} -- can be applied to topological pumping, yielding an order-by-order improvable (nearest-neighbor, third-neighbor, \dots) local Hamiltonian whose accuracy can be tuned against its range, including near the pump's gap-closing transition.

Real pumps are made of interacting particles. Interactions do not by themselves destroy
the quantization of the pumped charge, which survives as long as the many-body gap stays
open~\cite{niuQuantisedAdiabaticCharge1984}, but they can close that gap: an onsite Hubbard repulsion
breaks Thouless pumping in the spinful Rice-Mele
model~\cite{nakagawaBreakdownTopologicalThouless2018,bertokSplittingTopologicalCharge2022,walterQuantizationItsBreakdown2023}, and a competing nearest-neighbor repulsion
restores it~\cite{arguello-luengoStabilizationHubbardThoulessPumps2024}. In this regime the Bloch-vector construction of the
exact CD term is no longer available, and we find that the variational construction must
be adapted as well. The main result of this work is a local CD ansatz for the interacting
pump whose 14 coefficients are labeled by the interaction energy of each hop, together
with a ground-state-weighted variational action that determines them; it keeps the
interacting pump quantized at cycle times 25 to 50 times shorter than in
Ref.~\cite{arguello-luengoStabilizationHubbardThoulessPumps2024}.

This work examines the exact implementation, inherent limitations, and practical
approximations of CD driving in topological systems. Section~\ref{sec:adiabatic} reviews the theoretical
foundations of adiabatic and counterdiabatic dynamics and validates them on a driven
spin-1/2 system. Section~\ref{sec:topological} extends this framework to noninteracting topological matter:
the Kitaev chain shows that the range of the CD coupling diverges at a topological phase
transition, and the Rice-Mele Thouless pump shows that its exact CD term is non-local but
well approximated by a variational nearest-neighbor term. Section~\ref{sec:interacting}, the main part of
this work, turns to the interacting Rice-Mele pump, shows why the standard variational
construction fails there, and introduces an interaction-resolved ansatz that restores fast
quantized pumping. Section~\ref{sec:discussion} summarizes these findings and discusses open directions.

Throughout, we set $\hbar=1$ and measure energies in units of the hopping amplitude $t$ and times in units of $1/t$.

\section{Counterdiabatic Driving in Simple Quantum Systems}
\label{sec:adiabatic}

\subsection{Counterdiabatic Driving}
\label{subsec:cd}

An adiabatic evolution can be made exact, rather than merely approximate, by explicitly canceling the diabatic transitions it would otherwise induce. This is the idea behind CD driving: demanding that a state track an instantaneous eigenstate $|n(t)\rangle$ of the reference Hamiltonian $H_0(t)$ \emph{exactly}, i.e., $|\psi_n(t)\rangle = e^{i\zeta_n(t)}|n(t)\rangle$ with the phase fixed by Berry's result~\cite{berryQuantalPhaseFactors1984}, $\dot\zeta_n(t) = -E_n(t) + i\langle n(t)|\partial_t n(t)\rangle$ (the sum of a dynamical and a geometric contribution), singles out a unique total driving Hamiltonian $H(t) = H_0(t) + H_1(t)$ for which the Schr\"odinger equation is satisfied exactly. Requiring this for every eigenstate $n$ simultaneously fixes the auxiliary CD term to be

\begin{equation}
    H_1(t) = i\sum_n \Big( |\partial_t n\rangle\langle n| - \langle n | \partial_t n\rangle |n\rangle\langle n| \Big),
    \label{h1}
\end{equation}

with the full derivation given in Appendix~\ref{app:derivations} for completeness.

Equation~\eqref{h1} is directly related to the adiabatic gauge potential, $\mathcal{A}_\theta \equiv i\partial_\theta$, introduced in Ref.~\cite{kolodrubetzGeometryNonadiabaticResponse2017}. Writing $H_0$ in terms of a single time-dependent control parameter $\theta(t)$, so that $|\partial_t n\rangle = \dot\theta|\partial_\theta n\rangle$, Eq.~\eqref{h1} becomes
\begin{equation*}
H_1 = \dot\theta\Big(\mathcal{A}_\theta - \sum_n\langle n|\mathcal{A}_\theta|n\rangle\,|n\rangle\langle n|\Big),
\end{equation*}
i.e., the gauge potential with its diagonal (basis-dependent) part removed, scaled by the instantaneous evolution rate $\dot\theta$. This gauge-potential language recurs throughout the remainder of this work, most notably in Secs.~\ref{subsec:variational} and~\ref{sec:interacting}, where local approximations to $\mathcal{A}_\theta$ are constructed variationally for cases where the exact potential is not experimentally realizable.

\subsection{Application to Spin 1/2}
\label{subsec:spin12}

The general framework above is illustrated first with the simplest possible example: a spin-1/2 particle in a magnetic field,
\begin{equation}
    H = -\frac{1}{2}\mu\,\bm{\sigma}\cdot\vec{B} =
    -\frac{1}{2}\mu\begin{pmatrix}
        B_z&B_x-iB_y\\
        B_x+iB_y&-B_z
    \end{pmatrix},
\end{equation}
with $\bm{\sigma}$ the vector of Pauli matrices. Parameterizing the field direction by angles $\theta,\phi$ gives
\begin{equation}
    H_0 =
    -b\begin{pmatrix}
        \cos\theta&\sin{\theta}\, e^{-i\phi}\\
        \sin{\theta}\, e^{i\phi}&-\cos\theta
    \end{pmatrix}, \qquad b \equiv \mu B/2,
\end{equation}
with eigenvectors $|g\rangle = (\cos(\theta/2),\ e^{i\phi}\sin(\theta/2))^T$ and $|e\rangle = (\sin(\theta/2),\ -e^{i\phi}\cos(\theta/2))^T$.

\begin{figure}[tb]
    \centering
    \includegraphics[width=\columnwidth]{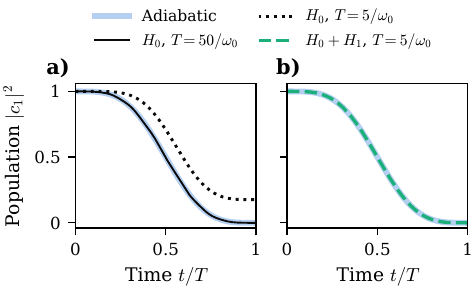}
    \caption{Population of state $|1\rangle$ of a spin-1/2 whose field rotates by $\pi$ in a
    time $T$, $\theta(t)=\pi(3s^2-2s^3)$ with $s=t/T$; the gap $\omega_0=2b$ sets the unit of
    time. The thick line is the instantaneous eigenstate. a)~With $H_0$ alone the system
    follows the eigenstate for a slow ramp, $T=50/\omega_0$, but not for a fast one,
    $T=5/\omega_0$. b)~Adding the CD term $H_1=(\dot\theta/2)\sigma_y$ makes the fast ramp
    follow the eigenstate exactly.}
    \label{fig:CDill}
\end{figure}

Consider the field rotating in the $xz$ plane, so that $\phi=0$ and only $\theta(t)$ varies in time. Equation~\eqref{h1} then gives the CD term
\begin{equation}
H_1(t) =
    \begin{pmatrix}
        0&-i\dot{\theta}/2\\
        i\dot{\theta}/2& 0
    \end{pmatrix}=
        \frac{\dot{\theta}}{2}\,\sigma_y .
\end{equation}
Figure~\ref{fig:CDill} confirms that this simple example already captures the essential feature of CD driving: with only $H_0$, the adiabatic approximation for the state populations visibly breaks down at fast rotation rates (large $\dot\theta$), whereas adding $H_1$ makes the exact and adiabatic solutions coincide for any $\dot\theta$~\cite{chenLewisRiesenfeldInvariantsTransitionless2011} -- the system tracks the ground state exactly regardless of driving speed.

Realizing $H_1$ directly is generally difficult experimentally, since it demands access to the full non-local, multi-qubit operator content of the auxiliary term. An established alternative is to synthesize an equivalent effective Hamiltonian using a fast periodic (Floquet) drive built only from local control fields, both for few-level systems~\cite{boyersFloquetengineeredQuantumState2019} and, via a systematic nested-commutator expansion, for many-body systems~\cite{claeysFloquetEngineeringCounterdiabaticProtocols2019}. This route is complementary to the variational, purely static approach developed below and is not pursued further here.

\section{Counterdiabatic Driving Applied to Topological Systems}
\label{sec:topological}
\subsection{Kitaev Chain}
\label{subsec:kitaev}

The Kitaev chain~\cite{kitaevUnpairedMajoranaFermions2001} is a paradigmatic toy model for a one-dimensional topological superconductor, describing a chain of spinless fermions with $p$-wave pairing. In real space (site basis), the Hamiltonian is
\begin{multline}
    H = \sum_{j=1}^{N-1} \left( -t(c_j^\dagger c_{j+1} + c_{j+1}^\dagger c_j) + \Delta(c_j c_{j+1} + c_{j+1}^\dagger c_j^\dagger) \right) \\
    -\mu \sum_{j=1}^N c_j^\dagger c_j,
    \nonumber
\end{multline}
where $\Delta$ is the mean-field pairing amplitude. Because the fermions are spinless, Pauli exclusion forbids two of them from occupying the same site, ruling out conventional on-site ($s$-wave) pairing. Fourier transforming to momentum space, $c_j = \frac{1}{\sqrt{N}}\sum_k e^{ikj}c_k$, and writing $H=\frac12\sum_k\Psi_k^\dagger H_k\Psi_k$ (up to a constant) with the Nambu spinor $\Psi_k=(c_k,c_{-k}^\dagger)^T$, gives the momentum-space Hamiltonian
\begin{equation}
    H_k =(-\mu-2t\cos k)\sigma_z+2\Delta \sin (k)\sigma_y = \vec{d}\cdot\bm{\sigma},
\end{equation}
with Bloch vector $\vec d = (0,\ 2\Delta \sin (k),\ -\mu-2t\cos k)$.

\begin{figure}[tb]
    \centering
    \includegraphics[width=\columnwidth]{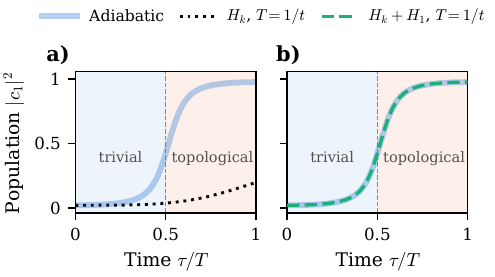}
    \caption{Population of $|1\rangle$ for the momentum mode $k=0.1\pi$ of the Kitaev chain
    ($\Delta=t$) as $\mu$ is ramped from $-4t$ to $0$ in a time $T=1/t$; the ramp crosses the
    topological transition, $\mu=-2t$, at $\tau=T/2$ (dashed line). The thick line is the
    instantaneous ground state. a)~Breakdown of the adiabatic approximation with $H_k$ alone.
    b)~Tracking of the ground state across the transition with $H_k+H_1$.}
    \label{fig:kitaevcd}
\end{figure}

This two-band form allows the CD correction to be obtained directly from the general two-level result (derived in Appendix~\ref{app:derivations}),
\begin{equation}
    H_1(k, \tau) = \frac{1}{2 |\vec{d}|^2} (\vec{d} \times \partial_\tau \vec{d}) \cdot \bm{\sigma},
    \label{CDterm}
\end{equation}
evaluated for a chemical potential driven in time as $\mu(\tau)$ (we use $\tau$ for the drive time here, to avoid confusion with the hopping amplitude $t$). This gives the explicit momentum-space correction
\begin{equation}
    H_1(k, \tau) = -\frac{\Delta \dot{\mu} \sin(k)}{(2t\cos(k) + \mu(\tau))^2 + 4\Delta^2\sin^2(k)} \sigma_x.
\end{equation}
Driving $\mu$ rapidly from $-4t$ to $0$, so that the system crosses the topological phase transition at $\mu=-2t$ (the chain is topological for $|\mu|<2t$, where $\vec d(k)$ winds once around the origin as $k$ runs through the Brillouin zone~\cite{kitaevUnpairedMajoranaFermions2001}), confirms that $H_1$ achieves transitionless driving across the transition itself: adding $H_1$ to $H_0$ reproduces the adiabatic populations exactly even though the gap closes at the critical point (Fig.~\ref{fig:kitaevcd}).

\begin{figure}[tb]
    \centering
    \includegraphics[width=\columnwidth]{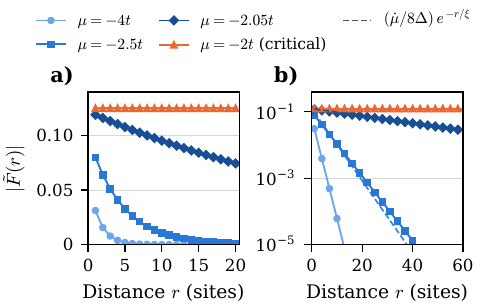}
    \caption{Exact CD coupling $\tilde F(r)$ between sites a distance $r$ apart, for
    $t=\Delta=\dot\mu=1$: deep in the trivial phase ($\mu=-4t$), approaching the transition
    ($\mu=-2.5t$ and $-2.05t$), and at the transition ($\mu=-2t$). a)~Linear scale.
    b)~Logarithmic scale. Away from the transition $\tilde F$ decays exponentially; the dashed
    lines show the asymptotic form $(\dot\mu/8\Delta)\,e^{-r/\xi}$ with the correlation length
    $\xi$ of Eq.~\eqref{eq:xi}. At the transition the coupling does not decay at all.}
    \label{fig:cdcoupling}
\end{figure}

Because $H_1$ was derived in momentum space, it is not obviously local. Transforming back to real space gives
\begin{equation}
    H_1 = i \sum_{l=1}^N \sum_{r=1}^{N/2} \tilde{F}(r) (c_{l+r}^\dagger c_l^\dagger + c_{l+r} c_l),
    \label{eq:kitaev_realspace}
\end{equation}
with coupling
\begin{equation}
    \tilde{F}(r) =  -\frac{1}{N} \sum_k \frac{ \Delta \dot{\mu} \sin(k) \sin(kr)}{(2t\cos(k) + \mu)^2 + 4\Delta^2\sin^2(k)},
\end{equation}
consistent with the equivalent result obtained in Ref.~\cite{delcampoAssistedFiniteRateAdiabatic2012} for the quantum Ising model, which is related to the Kitaev chain by a Jordan-Wigner transformation. Equation~\eqref{eq:kitaev_realspace} shows that $\tilde F(r)$ is the coupling strength that must be engineered between sites $l$ and $l+r$; its dependence on $r$ is plotted in Fig.~\ref{fig:cdcoupling}.

The range of this coupling can be understood analytically near the transition. Approximating the sum over $k$ as an integral for large $N$, and expanding near the gap closing ($2t+\mu\equiv m \ll 1$, $k\ll1$), the integral can be evaluated using Cauchy's theorem,
$$\int_{-\infty}^{\infty} \frac{k \sin(kr)}{k^2 + \left(\frac{2t + \mu}{2\Delta}\right)^2} \, dk = \pi e^{-\frac{|2t + \mu|}{2\Delta} r},$$
giving $\tilde F(r) \approx -\frac{\dot\mu}{8\Delta}e^{-|2t+\mu|r/2\Delta}$ and correlation length
\begin{equation}
    \xi = \frac{2\Delta}{|2t + \mu|}
    \label{eq:xi}
\end{equation}
(an equivalent expression holds near $\mu=2t$, obtained by expanding around $k=-\pi$). As expected, $\xi$ diverges at the transition, $\mu\to-2t$: engineering the exact CD term would require coupling site 1 to site 2 with the same strength as site 1 to site 1000. Figure~\ref{fig:cdcoupling}(b) confirms the exponential decay and its asymptotic form; at $\mu=-2t$ the coupling saturates at $|\tilde F|=\dot\mu/8\Delta$ and does not decay at all. This reflects a fundamental obstruction in quantum control: exact transitionless driving across a continuous quantum phase transition cannot be achieved with local Hamiltonian terms alone.

\subsection{Thouless Pump in Rice-Mele Model}
\label{subsec:ricemele}

The Rice-Mele model~\cite{riceElementaryExcitationsLinearly1982} is a 1D chain with two sites, $A$ and $B$, per unit cell. An electron can hop between the two sites within a unit cell with amplitude $t_1$, and between neighboring sites across adjacent unit cells with amplitude $t_2$:
\begin{multline}
     H = \sum_j \left[ \Delta c_{j,A}^\dagger c_{j,A} - \Delta c_{j,B}^\dagger c_{j,B} + t_1(c_{j,A}^\dagger c_{j,B} + \text{h.c.})\right]
     \\+\sum_j \left[t_2(c_{j,B}^\dagger c_{j+1,A} + \text{h.c.})\right].
     \nonumber
\end{multline}
Fourier transforming to momentum space, the Hamiltonian becomes
\begin{equation}
    H(k) = \begin{pmatrix}
    \Delta & t_1 + t_2 e^{-ik} \\ t_1 + t_2 e^{ik} & -\Delta
    \end{pmatrix}.
    \nonumber
\end{equation}
The Rice-Mele model is the natural setting for illustrating a Thouless pump~\cite{thoulessQuantizationParticleTransport1983,citroThoulessPumpingTopology2023}: rather than keeping $t_1$, $t_2$, and $\Delta$ static, they are made to depend on a slowly varying pump angle $\theta(t)$ that runs from $0$ to $2\pi$ over one cycle,
\begin{equation}
    t_1 = t+\delta\sin\theta,\qquad t_2 = t-\delta\sin\theta,\qquad \Delta(\theta) = -\Delta_0\cos\theta.
    \label{eq:pump_trajectory}
\end{equation}
In parameter space, as $\theta$ runs from $0$ to $2\pi$, the point $(\delta\sin\theta,\Delta_0\cos\theta)$ traces a closed ellipse around the origin. Although the Hamiltonian returns exactly to its initial form, the electron is displaced by exactly one unit cell over the cycle -- a topological invariant given by the Chern number of the Bloch bands over the torus spanned by $k$ and $\theta$~\cite{thoulessQuantizationParticleTransport1983}. This quantization is topologically protected: as long as the trajectory completes a loop around the origin, exactly one electron is pumped per cycle, regardless of the path's detailed shape.

\begin{figure}[tb]
    \centering
    \includegraphics[width=\columnwidth]{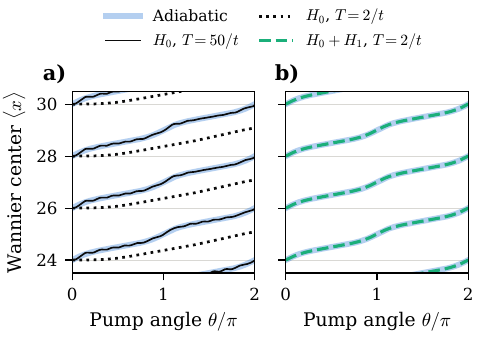}
    \caption{Wannier centers in the half-filled Rice-Mele chain (open chain of $N=30$ unit
    cells, $\delta=\Delta_0=0.5t$) over one pump cycle. The thick lines are the instantaneous
    ground state. a)~Without CD, a slow pump ($T=50/t$) follows the adiabatic path and moves
    every electron by one unit cell (two sites), while a fast pump ($T=2/t$) falls behind.
    b)~With the exact CD term, Eq.~\eqref{ricemeleH1}, the fast pump follows the adiabatic
    path.}
    \label{fig:ricemele}
\end{figure}

A Thouless pump is, however, an inherently adiabatic process and requires the parameters to be varied slowly. Simulating the pump using the real-space Hamiltonian shows that the quantized transport is preserved when the drive is slow, but breaks down when the pump is performed too quickly [Fig.~\ref{fig:ricemele}(a)].

Equation~\eqref{CDterm} showed that any $2\times2$ Bloch Hamiltonian has a CD term computable from the Bloch vector $\vec d$; here $\vec d = (t_1+t_2\cos k,\ t_2\sin k,\ \Delta)$. The complication specific to the Rice-Mele model is that all three parameters $\Delta, t_1, t_2$ vary with time, since a closed Thouless-pump cycle requires driving all of them, so $H_1$ includes three terms,
\begin{equation}
    H_1(\tau) = \dot{\Delta} \mathcal{A}_\Delta + \dot{t}_1 \mathcal{A}_{t_1} + \dot{t}_2 \mathcal{A}_{t_2},
\end{equation}
where $\mathcal{A}_\lambda$ is the gauge potential for the parameter $\lambda$. Writing $H_1(k) = f_x(k)\sigma_x + f_y(k)\sigma_y + f_z(k)\sigma_z$ and introducing the rotational velocity of the Hamiltonian in parameter space, $\Omega \equiv \vec{R} \times \dot{\vec{R}}$ with $\vec{R} = (t_1,t_2, \Delta)$, the coefficients are
\begin{gather*}
    f_x(k) = \frac{1}{2|\vec{d}|^2} \Omega_x \sin k,\qquad
    f_z(k) = \frac{1}{2|\vec{d}|^2} \Omega_z \sin k, \\
    f_y(k) = \frac{1}{2|\vec{d}|^2} \big(\Omega_y - \Omega_x \cos k\big).
\end{gather*}
With $H_1(k)$ in hand, transforming back to real space gives
\begin{multline}
    H_1(\tau) = i \sum_{l=1}^N \sum_r \Big[\tilde{J}_{AA}(r) (c_{l+r, A}^\dagger c_{l, A} - c_{l+r, B}^\dagger c_{l, B}) \Big] +\\
     i \sum_{l=1}^N \sum_r\Big[ \tilde{J}_{BA}(r) c_{l+r, A}^\dagger c_{l, B} + \text{h.c.}\Big],
     \label{ricemeleH1}
\end{multline}
with intra-sublattice and inter-sublattice couplings
\begin{gather*}
    \tilde{J}_{AA}(r) = \frac{\Omega_z}{4N} \sum_k \frac{\cos(k(r-1)) - \cos(k(r+1))}{|\vec{d}|^2}, \\
    \tilde{J}_{BA}(r) = \frac{1}{2N} \sum_k \frac{\Omega_x \cos(k(r-1)) - \Omega_y \cos(kr)}{|\vec{d}|^2}.
\end{gather*}
The $\tilde{J}_{AA}$ term requires electrons to hop from an $A$ site, over the intervening $B$ sublattice, to the $A$ site $r$ unit cells away -- a process with no counterpart in the original Hamiltonian, which only couples $A$ to $B$ sites. The sum over $r$ further indicates that the required hopping range extends across the entire lattice, spanning up to $N$ neighboring sites. Adding this exact CD term to the Thouless-pump evolution [Fig.~\ref{fig:ricemele}(b)] keeps the electron's motion perfectly adiabatic, as expected: it moves by exactly one unit cell per cycle, with the Wannier centers obtained numerically via the projected-position-operator method detailed in Appendix~\ref{app:numerics}.

\subsection{Variational Approach to CD: Rice-Mele Case}
\label{subsec:variational}

Implementing Eq.~\eqref{ricemeleH1} in an experiment is not feasible, owing to the infinite-range hopping it requires; a local approximation is therefore needed. Unlike the fine-tuned exact limit of Ref.~\cite{chielDynamicallyRobustCounterdiabatic2026} or the non-local realization of Ref.~\cite{liuShortcutsAdiabaticTopological2025}, we construct a systematically improvable local approximation with the variational approach of Sels and Polkovnikov~\cite{selsMinimizingIrreversibleLosses2017}. The CD term is written as a sum of simple, local operators, $H_1\approx\mathcal{A}=\sum_\mu\alpha_\mu O_\mu$, with coefficients chosen to minimize the action $S=\text{Tr}\,G^2$ of the generalized force $G=\dot H+i[\mathcal{A},H]$; Appendix~\ref{app:variational_derivation} shows that the minimum of $S$ over all operators is the exact CD term. The original Rice-Mele Hamiltonian is time-reversal invariant, and the CD gauge potential -- acting as a velocity operator in parameter space -- must be odd under time reversal, $T\mathcal{A}T^{-1}=-\mathcal{A}$. For spinless fermions $T$ is complex conjugation, so $\mathcal{A}$ must be purely imaginary in real space. Restricting to nearest-neighbor (NN) hoppings within a unit cell (intra-cell) and across adjacent unit cells (inter-cell) gives
\begin{equation*}
\mathcal{A}=\sum_j\Big[i\alpha_1\big(c^\dagger_{j,A}c_{j,B}-\text{h.c.}\big)+i\alpha_2\big(c^\dagger_{j,B}c_{j+1,A}-\text{h.c.}\big)\Big],
\end{equation*}
or, in momentum space,
\begin{equation}
    \mathcal{A}(k) = -\alpha_2 \sin(k)\sigma_x + \left(-\alpha_1 + \alpha_2 \cos(k)\right)\sigma_y.
    \label{eq:variational_ansatz}
\end{equation}
The generalized force, $G(k) = \vec{g}(k) \cdot \bm{\sigma}$, has components
\begin{align}
g_x(k)&=(\dot t_1+2\Delta\alpha_1)+(\dot t_2-2\Delta\alpha_2)\cos k,\nonumber\\
g_y(k)&=(\dot t_2-2\Delta\alpha_2)\sin k, \label{eq:gforce}\\
g_z(k)&=(\dot\Delta-2\alpha_1t_1+2\alpha_2t_2)+2(\alpha_2t_1-\alpha_1t_2)\cos k .\nonumber
\end{align}
The action
\begin{equation*}
    S = \frac{1}{2\pi}\int_{-\pi}^{\pi} \text{Tr}[G(k)^2] \, dk =\frac{1}{2\pi}\int dk\, (2|\vec{g}|^2)
\end{equation*}
can be integrated analytically, and $\partial S/\partial\alpha_i = 0$ gives $M\alpha = b$ for $\alpha=(\alpha_1,\alpha_2)$, with
\begin{gather}
M=\begin{pmatrix}2\Delta^2+2t_1^2+t_2^2 & -3t_1t_2\\ -3t_1t_2 & 2\Delta^2+t_1^2+2t_2^2\end{pmatrix},\nonumber\\
b=\begin{pmatrix}t_1\dot\Delta-\Delta\dot t_1\\ \Delta\dot t_2-t_2\dot\Delta\end{pmatrix},
\label{eq:Mb}
\end{gather}
so that $\alpha=M^{-1}b$.

\begin{figure}[tb]
    \centering
    \includegraphics[width=\columnwidth]{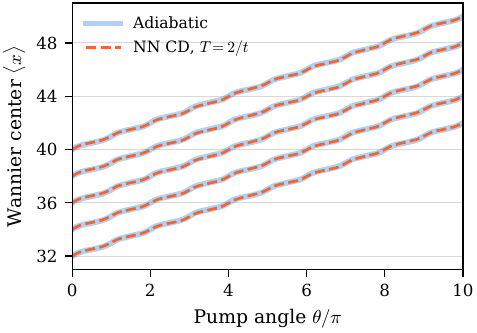}
    \caption{Wannier centers of five electrons on a periodic ring of $N=30$ unit cells over
    five pump cycles at $T=2/t$, with the NN variational CD term of Eq.~\eqref{eq:Mb}. The thick
    lines are the instantaneous ground state.}
    \label{fig:ricemelevar}
\end{figure}

Figure~\ref{fig:ricemelevar} shows the Thouless-pump evolution of the Wannier centers over five cycles with this NN term at $T=2/t$: it tracks the adiabatic path almost perfectly, with the Wannier centers within $0.05$ sites of the adiabatic path and fewer than $2\%$ of the electrons excited out of the lower band at any time. The pump trajectory, Eq.~\eqref{eq:pump_trajectory}, stays far from the gap closing at $t_1=t_2$, $\Delta=0$, so the exact CD couplings of Eq.~\eqref{ricemeleH1} decay rapidly with distance and the NN terms capture most of them.

\begin{figure}[tb]
    \centering
    \includegraphics[width=\columnwidth]{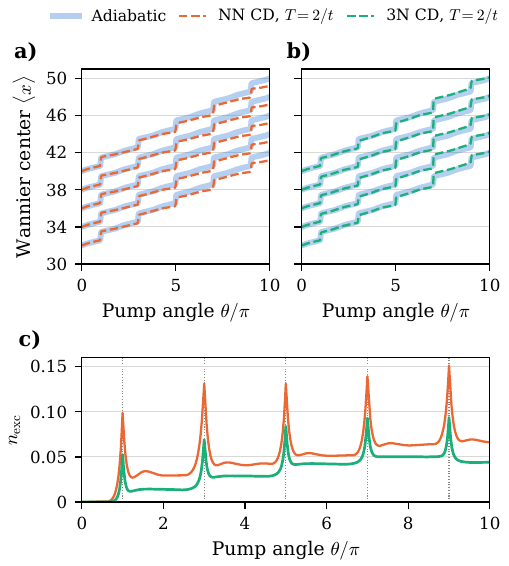}
    \caption{Pump path passing close to the gap closing, $c=0.95$ (smallest gap $0.05t$, at
    $t_1=t_2$, $\Delta=0.05\Delta_0$, reached at $\theta=\pi,3\pi,\dots$; dotted lines in c), on
    the same ring at $T=2/t$. a)~NN and b)~3N variational CD; the thick lines are the
    instantaneous ground state. c)~Fraction of electrons excited out of the instantaneous lower
    band, $n_\mathrm{exc}=1-\mathrm{Tr}(P_\mathrm{ad}P)/N$, where $P$ and $P_\mathrm{ad}$ project
    onto the occupied states of the evolved state and of the instantaneous ground state.}
    \label{fig:ricemelevar3N}
\end{figure}

Shifting the staggered potential, $\Delta=-\Delta_0(\cos\theta+c)$, brings the trajectory arbitrarily close to the gap closing, where long-range hoppings become significant. For $c=0.95$ [Fig.~\ref{fig:ricemelevar3N}(a)] the NN term still follows the adiabatic path, but excitations accumulate at each close approach to the gap closing. Extending the ansatz to all hoppings up to third neighbors (3N) -- the second-neighbor hopping within each sublattice and the two third-neighbor hoppings $(j,A)\leftrightarrow(j+1,B)$ and $(j,A)\leftrightarrow(j-2,B)$, five coefficients in total -- adds $\sin k\,\sigma_z$ and $\cos 2k$, $\sin 2k$ harmonics to Eq.~\eqref{eq:variational_ansatz} and reduces the fraction of excited electrons by about a third, from $6.6\%$ to $4.4\%$ after five cycles [Fig.~\ref{fig:ricemelevar3N}(b,c)].

For noninteracting fermions, a few local variational terms are therefore sufficient to approximate the exact, infinite-range CD protocol, even close to the gap closing. As we show next, this conclusion does not carry over to interacting fermions.

\section{Variational Approach to the Interacting Rice-Mele Model}
\label{sec:interacting}

The pumps of Sec.~\ref{sec:topological} are built from noninteracting fermions. In any realization --
electrons in a solid, ultracold atoms, or qubits -- the particles interact, and the
quantization of the pumped charge becomes a many-body property: it survives as long
as the many-body gap stays open along the whole cycle~\cite{niuQuantisedAdiabaticCharge1984}.
Interactions can, however, close that gap. A repulsive onsite Hubbard interaction $U$
competes with the staggered potential, which favors doubly occupied sites; once $U$ is
large enough, the many-body gap closes along the cycle and quantized pumping
breaks down~\cite{nakagawaBreakdownTopologicalThouless2018,bertokSplittingTopologicalCharge2022}, as observed with ultracold
fermions~\cite{walterQuantizationItsBreakdown2023}. Arg\"uello-Luengo \textit{et al.}~\cite{arguello-luengoStabilizationHubbardThoulessPumps2024}
showed that an additional nearest-neighbor repulsion $V$ restores quantized pumping in a
window around $V\approx U/2$: there $U$ suppresses the charge-density-wave instability
caused by $V$, $V$ suppresses the Mott instability caused by $U$, and the parent extended
Hubbard chain hosts a spontaneously dimerized bond-order wave. Their pumps were simulated
with cycle times $T=50/t$, deep in the adiabatic regime; faster cycles would reduce the
exposure of an experiment to heating and decoherence.

In this section we ask whether CD driving can run the same interacting pump more than an
order of magnitude faster, $T=2/t$ and $T=1/t$. Two ingredients of Sec.~\ref{sec:topological} no longer
apply: the exact gauge potential cannot be obtained from a Bloch vector [Eq.~\eqref{CDterm}], and,
as we show, the variational action of Sels and Polkovnikov~\cite{selsMinimizingIrreversibleLosses2017} no longer
produces a useful local CD term. We identify the reason, and construct a local ansatz
with 14 coefficients that keeps the interacting pump quantized at these speeds.

\subsection{Model and Numerical Method}
\label{sec:int_model}

We consider the spinful Rice-Mele chain with onsite and nearest-neighbor repulsion,
\begin{align}
H(\theta) ={}& \sum_{j,\sigma}\Big[t_1\big(c^\dagger_{j,A,\sigma}c_{j,B,\sigma}+\mathrm{h.c.}\big)\nonumber\\
 &\qquad+ t_2\big(c^\dagger_{j,B,\sigma}c_{j+1,A,\sigma}+\mathrm{h.c.}\big)\Big] \nonumber\\
 &+ \Delta\sum_{j}\big(n_{j,A}-n_{j,B}\big) \nonumber\\
 &+ U\sum_i n_{i\uparrow}n_{i\downarrow} + V\sum_i n_i n_{i+1},
 \label{eq:H_int}
\end{align}
where $\sigma\in\{\uparrow,\downarrow\}$, the index $i$ runs over all sites of the chain,
$n_i=n_{i\uparrow}+n_{i\downarrow}$, and $t_1$, $t_2$, $\Delta$ follow the pump path of
Eq.~\eqref{eq:pump_trajectory}. We take $\delta=0.9t$, $\Delta_0=0.5t$, $U=8t$ and $V=4t$. Up to a gauge
transformation that reverses the sign of all hoppings and a shift of the pump angle,
this point lies inside the window of stable quantized pumping found in
Ref.~\cite{arguello-luengoStabilizationHubbardThoulessPumps2024} ($3.9t\lesssim V\lesssim 4.3t$ at $U=8t$). We work at half
filling, with $N_\uparrow=N_\downarrow=L/2$ electrons on a ring of $L$ sites. Since
$H$ conserves $N_\uparrow$ and $N_\downarrow$ and treats both spins alike, an ideal cycle
moves one $\uparrow$ and one $\downarrow$ electron per unit cell by one unit cell: the
pumped charge per cycle is 2.

We solve the dynamics by exact diagonalization on periodic rings of $L=8$ and $L=12$
sites, i.e., in Hilbert spaces of $70^2=4900$ and $924^2=853\,776$ states. The
Hamiltonian is taken piecewise constant over 200 steps per cycle, and the exponential of
each step is applied to the state exactly (details and convergence checks are given in
Appendix~\ref{app:num_int}). We monitor two quantities. The pumped charge is
obtained from the many-body polarization of Resta~\cite{restaQuantumMechanicalPositionOperator1998},
\begin{equation}
Q(t) = \frac{1}{2\pi}\,\mathrm{Im}\ln\langle\psi(t)|\,e^{2\pi i\hat X/L}\,|\psi(t)\rangle,
\qquad \hat X=\sum_i x_i\,n_i ,
\label{eq:resta}
\end{equation}
where $x_i\in\{0,1,\dots,L-1\}$ is the position of site $i$. We follow $Q(t)$
continuously in time (without taking the phase modulo $2\pi$); it grows by exactly 2 per ideal cycle. The fidelity
\begin{equation}
\mathcal{F}(t)=\big|\langle\psi_0(\theta(t))|\psi(t)\rangle\big|^2
\label{eq:fidelity}
\end{equation}
measures the overlap with the instantaneous ground state $|\psi_0(\theta)\rangle$ of
Eq.~\eqref{eq:H_int}. In an ideal pump the system is in $|\psi_0(\theta)\rangle$ at every
moment; the electrons move because $|\psi_0(\theta)\rangle$ itself moves along the cycle,
not because they are excited. $\mathcal F$ is therefore the probability that the pump has
followed the ideal path. Because a single local excitation anywhere in the ring reduces
$\mathcal F$, the fidelity decreases with $L$ even at fixed accuracy per site, and we also
quote $\mathcal F^{1/L}$.

The smallest many-body gap along the cycle is $1.78t$ for $L=8$. At $T=2/t$, i.e.,
$\dot\theta=\pi t$, the drive is far from adiabatic: without CD the fidelity falls to
$\sim10^{-2}$ within the first cycle and the pumped charge stalls (Fig.~\ref{fig:int_2vs14}
and Table~\ref{tab:int_results}).

\subsection{Nearest-Neighbor Ansatz and the Infinite-Temperature Action}
\label{sec:int_NN}

The natural first attempt is to carry the nearest-neighbor (NN) ansatz of Sec.~\ref{subsec:variational} over
to both spin species. The Hamiltonian~\eqref{eq:H_int} is real and spin independent, so
the time-reversal argument of Sec.~\ref{subsec:variational} still requires a purely imaginary gauge
potential, and spin symmetry requires the same coefficients for both spins:
\begin{align}
\mathcal{A}^{\mathrm{NN}} = \sum_{j,\sigma}\Big[&i\alpha_1\big(c^\dagger_{j,A,\sigma}c_{j,B,\sigma}-\mathrm{h.c.}\big)\nonumber\\
 +{}&i\alpha_2\big(c^\dagger_{j,B,\sigma}c_{j+1,A,\sigma}-\mathrm{h.c.}\big)\Big].
\label{eq:A_NN_spin}
\end{align}
As in Sec.~\ref{subsec:variational}, the coefficients absorb the rate $\dot\theta$, so that the variational
operator approximates the CD term $H_1=\dot\theta\mathcal A_\theta$ itself, and the
generalized force is $G=\dot H+i[\mathcal A,H]$ with $\dot H=\dot\theta\,\partial_\theta H$.
It splits into $G=G_0+G_{\mathrm{int}}$, where $G_0=\dot H+i[\mathcal A,H_0]$ is the
noninteracting force of Sec.~\ref{subsec:variational} for each spin ($U$ and $V$ do not depend on time) and
$G_{\mathrm{int}}=i[\mathcal A,H_{\mathrm{int}}]$ contains the $U$ and $V$ terms. The interaction part follows from one identity. Let
$X=c^\dagger_{a\sigma}c_{b\sigma}$ move a spin-$\sigma$ electron from site $b$ to the
neighboring site $a$. Acting on any occupation-number configuration, $H_{\mathrm{int}}$ is
a number before the hop and another number after it, so
\begin{gather}
[X,H_{\mathrm{int}}]=-\,\Delta E\,X,\nonumber\\
\Delta E=(U-V)\big(n_{a\bar\sigma}-n_{b\bar\sigma}\big)+V\big(n_{a'}-n_{b'}\big),
\label{eq:dE}
\end{gather}
where $\Delta E$ is the change in interaction energy caused by the hop, $\bar\sigma$ is
the opposite spin, and $a'$, $b'$ are the outer neighbors of $a$ and $b$. The first term
collects the change on the two sites of the bond (a doublon formed or broken, together
with the bond between them), the second the change on the two outer bonds. $\Delta E$ only
reads the occupations of the four sites around the bond, which the hop does not change,
so it commutes with $X$. The reverse hop $X^\dagger$ costs $-\Delta E$, so
$[X^\dagger,H_{\mathrm{int}}]=\Delta E\,X^\dagger$, and each current $i(X-X^\dagger)$ in
$\mathcal A$ turns into a real hop weighted by its energy cost,
$i[\,i\alpha(X-X^\dagger),H_{\mathrm{int}}]=\alpha\,\Delta E\,(X+X^\dagger)$.

\begin{figure*}[t]
\centering
\includegraphics[width=\textwidth]{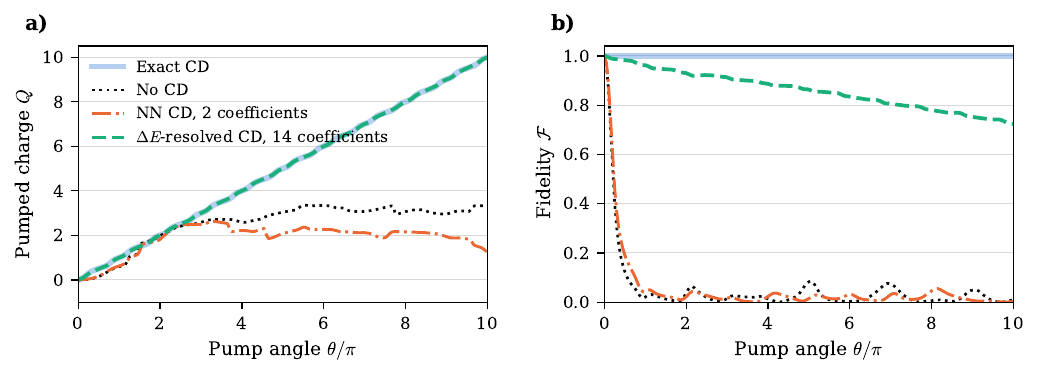}
\caption{Fast interacting Thouless pump on a ring of $L=8$ sites ($U=8t$, $V=4t$,
$\delta=0.9t$, $\Delta_0=0.5t$, $T=2/t$, five cycles). (a)~Pumped charge $Q$,
Eq.~\eqref{eq:resta}; an ideal pump transfers 2 per cycle. (b)~Fidelity $\mathcal F$
with the instantaneous ground state, Eq.~\eqref{eq:fidelity}. Without CD (dotted) the
pump fails within the first cycle. The NN ansatz with two coefficients obtained from the
infinite-temperature action, Eq.~\eqref{eq:Mc} (dash-dotted), does not improve on it.
The $\Delta E$-resolved ansatz with 14 coefficients, Eq.~\eqref{eq:A_dE} with
coefficients from Eq.~\eqref{eq:Mgs} (dashed), keeps $Q$ on the ideal line. The exact
CD term (thick solid) gives $\mathcal F=1$.}
\label{fig:int_2vs14}
\end{figure*}

The action of Sels and Polkovnikov is the Hilbert-Schmidt norm $\mathrm{Tr}\,G^2$, an
average over all configurations with equal weight -- an infinite-temperature average.
Evaluating it per unit cell (Appendix~\ref{app:interacting}), the cross term
$\mathrm{Tr}(G_0G_{\mathrm{int}})$ vanishes because $\Delta E$ averages to zero, and both
bond types give $\langle\Delta E^2\rangle=\tfrac12(U-V)^2+V^2$:
\begin{align}
\frac{S_\infty}{N} ={}& \frac12\,S_{\mathrm{free}}(\alpha_1,\alpha_2)\nonumber\\
 &+ \big(\alpha_1^2+\alpha_2^2\big)\Big[\tfrac12(U-V)^2+V^2\Big],
\label{eq:S_inf}
\end{align}
where $S_{\mathrm{free}}$ is the noninteracting action of Sec.~\ref{subsec:variational}. Its minimum is
\begin{equation}
\big(M+c\,\openone\big)\,\alpha=b,\qquad c=\tfrac14(U-V)^2+\tfrac12V^2,
\label{eq:Mc}
\end{equation}
with $M$ and $b$ from Eq.~\eqref{eq:Mb}. The interactions enter only as a positive shift of the
diagonal of $M$: they penalize the CD amplitudes without telling the ansatz how to react
to them. For $U=8t$ and $V=4t$, $c=12t^2$ exceeds all entries of $M$, and along the cycle
the coefficient vector $(\alpha_1,\alpha_2)$ is 2.7 to 25 times shorter than its
noninteracting value. The resulting CD term does not help at all
(Fig.~\ref{fig:int_2vs14}, dash-dotted lines): at $T=2/t$ the
fidelity is $0.02$ after the first cycle, no better than without CD, and the pumped charge
drifts away from the quantized value. A single coefficient per bond has to serve hops
whose interaction cost ranges from $-12t$ to $+12t$, and the infinite-temperature average
is dominated by high-energy configurations that the pump never visits.

\subsection{Exact Gauge Potential and a Ground-State Action}
\label{sec:int_exact}

To see what an interacting CD term has to contain, we return to the exact gauge potential.
The derivation of Appendix~\ref{app:derivations} does not assume a two-level system: for many-body eigenstates
$H|n\rangle=E_n|n\rangle$ it gives
\begin{equation}
\mathcal A_\theta = i\sum_{m\neq n}|m\rangle\,
\frac{\langle m|\partial_\theta H|n\rangle}{E_n-E_m}\,\langle n| .
\label{eq:AGP_exact}
\end{equation}
For $U=V=0$ the Hamiltonian is a sum of independent $2\times2$ Bloch blocks, and
Eq.~\eqref{eq:AGP_exact} reduces block by block to Eq.~\eqref{CDterm} (Appendix~\ref{app:interacting}). With
interactions there is no Bloch vector, and Eq.~\eqref{eq:AGP_exact} involves all
many-body eigenstates. The CD term is $H_1=\dot\theta\mathcal A_\theta$ (Sec.~\ref{subsec:cd}), and for
the pump only its action on the instantaneous ground state $|0\rangle$ matters,
\begin{equation}
|\eta\rangle\equiv\dot\theta\,\mathcal A_\theta|0\rangle
 = -i\,(H-E_0)^{-1}\,\mathcal Q\,\dot H|0\rangle,
\label{eq:eta}
\end{equation}
with $\mathcal Q=1-|0\rangle\langle0|$,
which follows from Eq.~\eqref{eq:AGP_exact} by writing $\mathcal Q=\sum_{n\neq0}|n\rangle\langle n|$ and
$(H-E_0)^{-1}|n\rangle=|n\rangle/(E_n-E_0)$. The projector $\mathcal Q$ removes the
ground-state component, on which $H-E_0$ cannot be inverted. In practice $|\eta\rangle$ is
obtained from the ground state with a single linear solve; no other eigenstates are needed.

For an approximate $\mathcal A$, the generalized force no longer vanishes off the
diagonal. Using $H|0\rangle=E_0|0\rangle$ and the fact that the exact CD term makes
$\langle n|G|0\rangle$ vanish (Appendix~\ref{app:interacting}),
\begin{equation}
\langle n|G|0\rangle = i(E_0-E_n)\big[\langle n|\mathcal A|0\rangle-\langle n|\eta\rangle\big],
\qquad n\neq0 .
\label{eq:Gn0}
\end{equation}
The quantity that drives transitions out of the ground state is the residual coupling
$\langle n|\mathcal A|0\rangle-\langle n|\eta\rangle$ between $|0\rangle$ and $|n\rangle$
that remains in the frame co-moving with the eigenstates. This suggests the ground-state
action
\begin{equation}
S_{\mathrm{gs}}[\mathcal A]=\sum_{n\neq0}\frac{|\langle n|G|0\rangle|^2}{(E_n-E_0)^2}
 = \big\|\,\mathcal Q\,\mathcal A|0\rangle-|\eta\rangle\big\|^2 .
\label{eq:S_gs}
\end{equation}
For $\mathcal A=\sum_\mu\alpha_\mu O_\mu$ it is a linear least-squares fit of $|\eta\rangle$
by the vectors $\mathcal Q O_\mu|0\rangle$, minimized by
\begin{gather}
\sum_\nu M^{\mathrm{gs}}_{\mu\nu}\alpha_\nu=b^{\mathrm{gs}}_\mu,\nonumber\\
M^{\mathrm{gs}}_{\mu\nu}=\mathrm{Re}\,\langle0|O_\mu\mathcal Q O_\nu|0\rangle,\qquad
b^{\mathrm{gs}}_\mu=\mathrm{Re}\,\langle0|O_\mu\mathcal Q|\eta\rangle .
\label{eq:Mgs}
\end{gather}
Both actions vanish for the exact gauge potential; they differ in which errors an
incomplete ansatz is allowed to make. $S_\infty$ weights all states
equally, whereas $S_{\mathrm{gs}}$ keeps only the ground-state column of $G$, weighted
towards the low-energy states into which leakage is most harmful. $S_{\mathrm{gs}}$
corresponds to the limit of the weighted actions of Ohga and Hatomura~\cite{ohgaImprovingVariationalCounterdiabatic2026} in
which only the ground-state column of the deviation from the exact gauge potential is
penalized; here we evaluate it directly from $|0\rangle$ and $|\eta\rangle$. Since
$|\eta\rangle$ is proportional to $\dot\theta$ and $|0\rangle$ depends only on $\theta$,
the coefficients take the form $\alpha_\mu(t)=\dot\theta\,\tilde\alpha_\mu(\theta)$: they
are computed once on a grid of $\theta$ (40 points per cycle, interpolated by periodic
splines) and reused for every period $T$.

With $S_{\mathrm{gs}}$ the two NN coefficients do better, but the pump still fails after a
few cycles ($\mathcal F=0.59$, $0.11$, $0.11$ after cycles 1, 3 and 5 at $T=2/t$;
Table~\ref{tab:int_results}). The limitation is now the operator content: along the cycle
the two NN currents capture 83--93\% of $\|\eta\|^2$. Adding one-body currents of arbitrary
range hardly helps: on the 8-site ring the optimal one-body ansatz captures at most 3\% more
of $\|\eta\|^2$ than the two NN currents. In particular, no single-particle CD term, such
as one obtained by inserting a mean-field Bloch vector into Eq.~\eqref{CDterm}, can capture the
missing part.

\subsection{Interaction-Resolved Ansatz}
\label{sec:int_dE}

Equation~\eqref{eq:AGP_exact} indicates what is missing. Since $\partial_\theta H$ contains
only the hoppings and the staggered potential, the numerator $\langle m|\partial_\theta
H|n\rangle$ connects configurations that differ by a single NN hop. With interactions,
the energy denominator depends on the interaction energy $\Delta E$, Eq.~\eqref{eq:dE},
that this hop costs, which ranges from $-12t$ to $12t$ for our parameters, well beyond the
hopping scale. The exact gauge potential is thus approximately a
NN current whose amplitude depends on $\Delta E$ -- precisely what a single coefficient per
bond cannot represent, and precisely the structure that $G_{\mathrm{int}}$ penalized in
Sec.~\ref{sec:int_NN}. We therefore resolve every NN current by the interaction energy of
its hop:
\begin{align}
\mathcal A^{\Delta E}=\sum_{\varepsilon}\Big[&\alpha^{\mathrm{intra}}_{\varepsilon}\sum_{j,\sigma}J^{\mathrm{intra}}_{j\sigma}P^{\mathrm{intra}}_{j\sigma,\varepsilon}\nonumber\\
 +{}&\alpha^{\mathrm{inter}}_{\varepsilon}\sum_{j,\sigma}J^{\mathrm{inter}}_{j\sigma}P^{\mathrm{inter}}_{j\sigma,\varepsilon}\Big],
\label{eq:A_dE}
\end{align}
with the NN currents
$J^{\mathrm{intra}}_{j\sigma}=i(c^\dagger_{j,A,\sigma}c_{j,B,\sigma}-\mathrm{h.c.})$ and
$J^{\mathrm{inter}}_{j\sigma}=i(c^\dagger_{j,B,\sigma}c_{j+1,A,\sigma}-\mathrm{h.c.})$.
$P^{\mathrm{intra}}_{j\sigma,\varepsilon}$ is the projector onto configurations in which
the hop of a spin-$\sigma$ electron from $(j,B)$ to $(j,A)$ changes the interaction energy
by $\Delta E=\varepsilon$, Eq.~\eqref{eq:dE} with $a=(j,A)$, $b=(j,B)$, $a'=(j-1,B)$,
$b'=(j+1,A)$; $P^{\mathrm{inter}}_{j\sigma,\varepsilon}$ is defined in the same way for the
hop from $(j+1,A)$ to $(j,B)$. The projectors depend only on the occupations of the four
sites around the bond, which the hop leaves unchanged, so every term in
Eq.~\eqref{eq:A_dE} is Hermitian, odd under time reversal, and supported on four
consecutive sites. The label $\varepsilon$ is not a weight: acting on a configuration,
$\mathcal A^{\Delta E}$ performs every NN hop with the coefficient of that hop's energy
class, and the variational fit determines how the amplitude depends on $\varepsilon$.

In Eq.~\eqref{eq:dE} the first bracket takes three values and the second five, so for
generic $U$ and $V$ there are 15 classes per bond type. For $U=2V$ the 15 combinations
fall onto seven distinct energies, $\varepsilon\in\{0,\pm V,\pm2V,\pm3V\}$, i.e.,
$\{0,\pm4t,\pm8t,\pm12t\}$ here
(Table~\ref{tab:dE_classes} in Appendix~\ref{app:interacting}), leaving 14 coefficients. They are not all
independent functions of $\theta$: translation by one site maps $H(\theta)$ onto
$H(\theta+\pi)$ and intracell onto intercell bonds, so
$\alpha^{\mathrm{inter}}_\varepsilon(\theta)=\alpha^{\mathrm{intra}}_\varepsilon(\theta+\pi)$,
which our fits reproduce to $10^{-8}$. For $U=V=0$ every hop has $\varepsilon=0$ and
Eq.~\eqref{eq:A_dE} reduces to the NN ansatz~\eqref{eq:A_NN_spin}. The coefficients are
obtained from $S_{\mathrm{gs}}$, Eq.~\eqref{eq:Mgs}.

The fit leaves 1.7--4.6\% of $\|\eta\|^2$ unexplained along the cycle for $L=8$, and
1.3--4.9\% for $L=12$. For comparison, the complete set of 168 independent
spin-symmetric, time-reversal-odd operators supported on three consecutive sites leaves
about 2\% ($L=8$). Within that set almost all of the weight is carried by
single-particle NN hops conditioned on the surrounding occupations, which is the class
spanned by Eq.~\eqref{eq:A_dE}; pair hopping and longer-range hops contribute less than 1\%.

\subsection{Fast Interacting Pumps}
\label{sec:int_results}

Figure~\ref{fig:int_2vs14} compares the protocols at $T=2/t$ on the 8-site ring. With the
$\Delta E$-resolved CD term the pumped charge follows the ideal line, $Q=10.00$ after five
cycles, and the fidelity is $0.93$ after the first cycle, decreasing slowly to $0.72$ after
five. The exact CD term restricted to the ground state, the rank-two operator
$|\eta\rangle\langle0|+|0\rangle\langle\eta|$, gives $\mathcal F=1$ by construction; it is
strongly non-local and serves only as a benchmark.

\begin{figure*}[t]
\centering
\includegraphics[width=\textwidth]{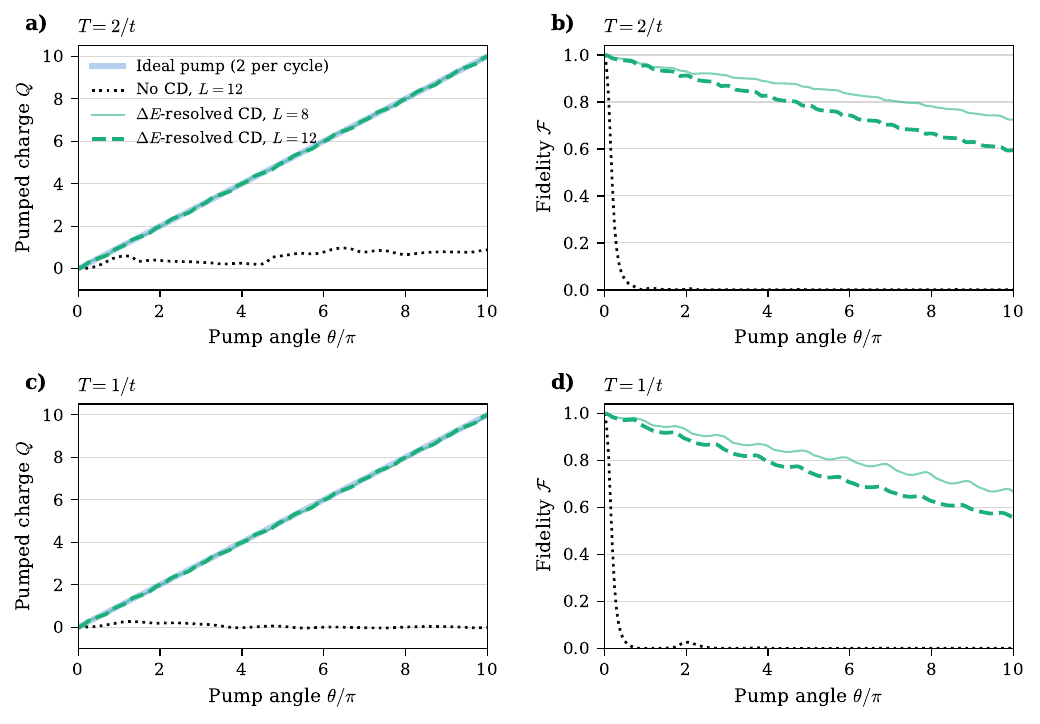}
\caption{$\Delta E$-resolved CD for rings of $L=8$ (thin solid) and $L=12$ sites (dashed),
at $T=2/t$ (a,b) and $T=1/t$ (c,d); other parameters as in Fig.~\ref{fig:int_2vs14}.
(a,c)~Pumped charge, compared with the ideal value of 2 per cycle (thick line).
(b,d)~Fidelity with the instantaneous ground state. Without CD (dotted, $L=12$) the pump
fails within the first cycle at both speeds.}
\label{fig:int_sizes}
\end{figure*}

Figure~\ref{fig:int_sizes} and Table~\ref{tab:int_results} extend the comparison to $L=12$
and to $T=1/t$. The pumped charge stays quantized at both sizes and both speeds
($Q=9.98$--$10.00$ after five cycles), while without CD the pump fails within the first
cycle. The fidelity is lower for $L=12$, as expected for a many-body overlap, but the
fidelity per site after one cycle at $T=2/t$ is $\mathcal F^{1/L}=0.991$ for $L=8$ and
$0.992$ for $L=12$: the accuracy per site does not degrade with system size. Halving the
period from $2/t$ to $1/t$ lowers the first-cycle fidelity only from $0.91$ to $0.89$
($L=12$). Both the CD term and its residual error scale with $\dot\theta$, while the
duration of a cycle scales with $1/\dot\theta$, so the loss per cycle is nearly
independent of the speed -- in sharp contrast with the uncorrected pump.

The fitted coefficients are shown in Appendix~\ref{app:coeffs}
(Fig.~\ref{fig:int_coeffs}). They vary smoothly along the cycle and are largest for the
classes with small $|\varepsilon|$, as expected from the energy denominators of
Eq.~\eqref{eq:AGP_exact}.

The remaining loss of fidelity per cycle, 5--7\% for $L=8$ and about 10\% for $L=12$,
comes from the few percent of $|\eta\rangle$ that Eq.~\eqref{eq:A_dE} does not capture.
It can be reduced further by enlarging the ansatz: with all 168 three-site operators the
fidelity at $T=2/t$ and $L=8$ is $0.97$, $0.92$ and $0.87$ after cycles 1, 3 and 5. The
$\Delta E$-resolved ansatz is a compact compromise, with 14 instead of 168 coefficients and
a clear interpretation in terms of the energy denominators of the exact gauge potential.

The operators in Eq.~\eqref{eq:A_dE} are density-dependent hoppings acting on four
consecutive sites. On digital quantum simulators each of them decomposes into a few
multi-qubit gates per bond; in optical lattices, density-dependent tunneling has been
engineered with periodically modulated lattices~\cite{gorgRealizationDensitydependentPeierls2019}. The coefficients require
the instantaneous ground state and one linear solve per value of $\theta$, a classical
precomputation that can be carried out for long chains with matrix-product-state methods.

\begin{table}[t]
\caption{End-of-cycle fidelity $\mathcal F$ after cycles 1, 3 and 5, and pumped charge $Q$
after five cycles (ideal value 10), for $U=8t$, $V=4t$. The coefficients of the CD terms
come from the infinite-temperature action $S_\infty$, Eq.~\eqref{eq:Mc}, or from the
ground-state action $S_{\mathrm{gs}}$, Eq.~\eqref{eq:Mgs}.}
\label{tab:int_results}
\begin{ruledtabular}
\begin{tabular}{llcccc}
$L$, $T$ & Protocol & $\mathcal F_1$ & $\mathcal F_3$ & $\mathcal F_5$ & $Q_5$ \\
\hline
8, $2/t$ & No CD & 0.037 & 0.003 & 0.010 & 3.31 \\
 & NN (2 coeff.), $S_\infty$ & 0.022 & 0.031 & 0.003 & 1.24 \\
 & NN (2 coeff.), $S_{\mathrm{gs}}$ & 0.59 & 0.11 & 0.11 & 6.99 \\
 & $\Delta E$-resolved (14), $S_{\mathrm{gs}}$ & 0.93 & 0.83 & 0.72 & 10.00 \\
 & 3-site (168), $S_{\mathrm{gs}}$ & 0.97 & 0.92 & 0.87 & 10.00 \\
 & Exact CD & 1 & 1 & 1 & 10.00 \\
8, $1/t$ & No CD & 0.086 & 0.082 & 0.057 & $-0.02$ \\
 & $\Delta E$-resolved (14), $S_{\mathrm{gs}}$ & 0.93 & 0.81 & 0.67 & 9.98 \\
12, $2/t$ & No CD & 0.003 & $<10^{-3}$ & $<10^{-3}$ & 0.89 \\
 & $\Delta E$-resolved (14), $S_{\mathrm{gs}}$ & 0.91 & 0.74 & 0.59 & 10.00 \\
12, $1/t$ & No CD & 0.027 & $<10^{-3}$ & $<10^{-3}$ & $-0.01$ \\
 & $\Delta E$-resolved (14), $S_{\mathrm{gs}}$ & 0.89 & 0.71 & 0.56 & 9.99 \\
\end{tabular}
\end{ruledtabular}
\end{table}

\section{Discussion and Outlook}
\label{sec:discussion}

This work has examined the limitations and practical implementation of CD driving in
topological matter. For noninteracting fermions, the Kitaev chain and the Rice-Mele model
show that exact transitionless driving across a topological phase transition, or around a
pump cycle, requires couplings whose range grows without bound near the gap closing.
Away from it, the variational framework of Sels and Polkovnikov yields local
nearest-neighbor approximations that keep the Rice-Mele pump on its adiabatic path at
cycle times far below the adiabatic limit.

The main result concerns interacting pumps. In the spinful Rice-Mele chain with onsite
repulsion $U=8t$ and nearest-neighbor repulsion $V=4t$, where the competition of the two
interactions stabilizes quantized pumping~\cite{arguello-luengoStabilizationHubbardThoulessPumps2024}, the standard
variational construction fails. The infinite-temperature action only penalizes the CD
amplitudes, and the resulting two-parameter nearest-neighbor term does not prevent the
breakdown of the pump at $T=2/t$. The exact many-body gauge potential explains why: its
amplitude for each hop depends on the interaction energy $\Delta E$ that the hop costs.
Resolving the nearest-neighbor currents by $\Delta E$ -- one coefficient per value of
$\Delta E$ and bond type, 14 in total for $U=2V$ -- and determining the coefficients with a
ground-state-weighted action restores quantized pumping at $T=2/t$ and $T=1/t$, 25 to 50
times faster than the protocols of Ref.~\cite{arguello-luengoStabilizationHubbardThoulessPumps2024}. The accuracy per site
is the same for rings of 8 and 12 sites, the fitted coefficients are similar for the two
sizes, and the loss of fidelity per cycle hardly depends on the speed. Two lessons
generalize beyond this model: for interacting systems the choice of the variational norm is
as important as the choice of operators, and the energy denominators of the exact gauge
potential suggest which operators to use -- here, hops conditioned on the surrounding
occupations.

Several directions remain open. Our results are limited to rings of up to 12 sites by
exact diagonalization. Both ingredients of the coefficients, the instantaneous ground state
and the solution of the linear system~\eqref{eq:eta}, are standard tasks for
matrix-product-state methods, so the approach can be extended to long chains; the
size-independent accuracy per site found for 8 and 12 sites suggests that it does not
degrade with system size. The remaining loss of fidelity per cycle can be reduced by enlarging
the operator set, for example with all three-site operators, or by combining the local CD
term with an optimized pump path. The interaction-resolved construction does not depend on
the details of the Rice-Mele model and should apply to other interacting adiabatic
protocols, such as ground-state preparation in Hubbard-type models. Finally, extending
variational CD driving to two-dimensional topological phases, such as the toric code,
would test whether local approximations can prepare topologically ordered ground states
at speeds relevant for topological quantum computation.

\acknowledgements
I express my sincere gratitude to Stewart Morawetz and Bernardo Barrera for many helpful discussions that shaped these ideas and helped make sense of the results. I also deeply thank my research supervisor, Anatoli Polkovnikov, for his exceptional guidance and inspiring ideas.

\appendix

\section{Derivation of the Counterdiabatic Term}
\label{app:derivations}

\subsection{General derivation of the counterdiabatic term}
\label{app:derivations_general}

Here we derive Eq.~\eqref{h1}, the CD term that makes a state track an instantaneous eigenstate $|n(t)\rangle$ of $H_0(t)$ exactly. The target state is $|\psi_n(t)\rangle = e^{i\zeta_n(t)}|n(t)\rangle$, with $\zeta_n(t)$ the total (dynamical plus geometric) phase fixed by Berry's result~\cite{berryQuantalPhaseFactors1984},
$$\dot{\zeta}_n(t) = -E_n(t) + i\langle n(t) | \partial_t n(t)\rangle.$$
Demanding that $|\psi_n(t)\rangle$ satisfy the time-dependent Schr\"odinger equation, $H(t) |\psi_n(t)\rangle = i\partial_t |\psi_n(t)\rangle$, and differentiating the ansatz gives
\begin{equation}
    i \partial_t |\psi_n(t)\rangle = i \left( i\dot{\zeta}_n e^{i\zeta_n} |n\rangle + e^{i\zeta_n} |\partial_t n\rangle \right).
\end{equation}
Substituting the expression for $\dot\zeta_n$ yields
$$i \partial_t |\psi_n(t)\rangle = \Big( E_n |n\rangle - i\langle n | \partial_t n\rangle |n\rangle + i |\partial_t n\rangle \Big) e^{i\zeta_n}.$$
Writing $H(t)$ in its own (as yet unknown) eigenbasis, $H(t) = \sum_n \big( H(t) |\psi_n(t)\rangle \big) \langle \psi_n(t) |$, and inserting the result above together with $\langle \psi_n(t) | = e^{-i\zeta_n} \langle n(t) |$, gives the total driving Hamiltonian $H(t)=H_0(t)+H_1(t)$, where $H_0$ contributes the $E_n|n\rangle\langle n|$ term and

$$H_1(t) = i\sum_n \Big( |\partial_t n\rangle\langle n| - \langle n | \partial_t n\rangle |n\rangle\langle n| \Big),$$

which recovers Eq.~\eqref{h1} of the main text.

\subsection{Counterdiabatic term for a two-level system}
\label{app:derivations_twolevel}

We now derive Eq.~\eqref{CDterm}, the CD correction written in terms of the Bloch vector $\vec d$. Starting from Eq.~\eqref{h1}, inserting the resolution of the identity $\sum_m|m\rangle\langle m| = \mathbbm{1}$, and noting that the $m=n$ terms cancel, $H_1$ can equivalently be written as a sum over $m\neq n$ only,
\begin{equation}
    H_1 = i\sum_{n} \sum_{m \neq n} |m\rangle \langle m | \dot{n} \rangle \langle n|.
\end{equation}
Taking the time derivative of the eigenvalue equation $H_0|n\rangle = E_n|n\rangle$ and rearranging gives
\begin{equation}
    \langle m | \dot{n}\rangle = \frac{\langle m | \dot{H}_0 |n\rangle}{E_n - E_m}.
\end{equation}
Substituting into the expression for $H_1$ and specializing to a two-level system with eigenvalues $E_\pm = \pm d$,
\begin{equation}
    H_1 = \frac{i}{2d} \Big( |-\rangle\langle -| \dot{H}_0 |+\rangle\langle +| - |+\rangle\langle +| \dot{H}_0 |-\rangle\langle -| \Big).
\end{equation}
Writing $H_0$ in its eigenbasis, $H_0 = d(|+\rangle\langle +| - |-\rangle\langle -|)$, gives the commutator
\begin{equation}
    [H_0, \dot{H}_0] = 2d \Big( |+\rangle\langle +| \dot{H}_0 |-\rangle\langle -| - |-\rangle\langle -| \dot{H}_0 |+\rangle\langle +| \Big),
\end{equation}
so that $H_1$ can be written compactly as
\begin{equation}
    H_1 =  -\frac{i}{4d^2} [H_0, \dot{H}_0].
\end{equation}
Applying the Pauli identity $[\vec{a}\cdot\vec{\sigma}, \vec{b}\cdot\vec{\sigma}] = 2i (\vec{a} \times \vec{b}) \cdot \vec{\sigma}$ then gives Eq.~\eqref{CDterm}.

\section{Equivalence of the Variational Principle and the Exact Adiabatic Gauge Potential}
\label{app:variational_derivation}

This appendix proves the equivalence between the exact commutator equation defining the adiabatic gauge potential and the variational principle used in Sec.~\ref{subsec:variational} to construct its local approximations.

Consider the off-diagonal matrix elements of the commutator $[\mathcal{A}_\lambda, H_0]$: $\langle m | [\mathcal{A}_\lambda, H_0] | n \rangle = (E_n - E_m) \langle m | \mathcal{A}_\lambda | n \rangle$. Substituting the definition of $\mathcal{A}_\lambda$ into this result,
$$\langle m | [\mathcal{A}_\lambda, H_0] | n \rangle = i(E_n - E_m) \langle m | \partial_\lambda n \rangle.$$
As shown in Appendix~\ref{app:derivations}, $\langle m | \partial_\lambda H_0 | n \rangle = (E_n-E_m) \langle m|\partial_\lambda n \rangle$, so combining the two relations gives
$$\langle m | \Big( \partial_\lambda H_0 + i[\mathcal{A}_\lambda, H_0] \Big) | n \rangle = 0 \quad \text{for all } m \neq n.$$
The operator in parentheses is the generalized force $G \equiv \partial_\lambda H_0 + i[\mathcal{A}_\lambda, H_0]$ of the main text, with the derivative taken with respect to $\lambda$ instead of time. All off-diagonal elements of $G$ vanish, so $G$ is diagonal in the eigenbasis of $H_0$ and therefore commutes with it:
$$\big[\partial_\lambda H_0 + i[\mathcal{A}_\lambda, H_0],\, H_0\big] = 0.$$

It remains to show that solving this commutator equation is equivalent to minimizing the squared trace, or action,
$$S(\mathcal{A}_\lambda) = \text{Tr}\,G^2 = \text{Tr}\Big[ \big(\partial_\lambda H_0 + i[\mathcal{A}_\lambda, H_0]\big)^2 \Big].$$
Introducing a small, arbitrary variation $\mathcal{A}_\lambda \to \mathcal{A}_\lambda + \delta \mathcal{A}_\lambda$ causes $G \to G + i[\delta \mathcal{A}_\lambda, H_0]$, so
$$S + \delta S = \text{Tr}\Big[ (G + i[\delta \mathcal{A}_\lambda, H_0])^2 \Big].$$
Keeping only the terms linear in $\delta\mathcal{A}_\lambda$ (the first variation $\delta S$) and using the cyclic property of the trace, $\text{Tr}[AB]=\text{Tr}[BA]$, the two linear terms are identical:
$$\delta S = 2i \text{Tr}\Big[ G [\delta \mathcal{A}_\lambda, H_0] \Big].$$
Expanding the commutator inside the trace and applying the cyclic property again to shift $H_0$ to the front of the first term,
\begin{widetext}
\begin{equation}
\delta S = 2i \Big( \text{Tr}[H_0 G \delta \mathcal{A}_\lambda] - \text{Tr}[G H_0 \delta \mathcal{A}_\lambda] \Big) = 2i \text{Tr}\Big[ (H_0 G - G H_0) \delta \mathcal{A}_\lambda \Big].
\end{equation}
\end{widetext}
\noindent This collapses back into a commutator,
$$\delta S = 2i \text{Tr}\Big[ [H_0, G] \delta \mathcal{A}_\lambda \Big].$$
At a minimum of $S(\mathcal{A}_\lambda)$, $\delta S$ must vanish for any variation $\delta\mathcal{A}_\lambda$, which is only possible if the operator multiplying it is identically zero:
$$[H_0, G] = 0.$$
Conversely, since $S$ is a non-negative quadratic function of $\mathcal{A}_\lambda$, every such stationary point is a global minimum. This is exactly the commutator equation derived above, confirming that the variational minimization used in Sec.~\ref{subsec:variational} and the exact adiabatic gauge potential condition are equivalent.

\section{Numerical Methods}
\label{app:numerics}

In all simulations the Hamiltonian is held constant during each of the 200 time steps per pump
cycle, and the exponential of each step is applied exactly.

\subsection{Noninteracting chains (Figs.~\ref{fig:ricemele}--\ref{fig:ricemelevar3N})}
\label{app:num_free}

Without interactions, $H_0$ and all CD terms of Sec.~\ref{sec:topological} are quadratic, so the
state remains a Slater determinant. It is stored as the $L\times N$ matrix $V$ whose columns are
the $N$ occupied orbitals, starting from the lowest eigenvectors of $h_0(\theta=0)$ and updated as
$V\to e^{-ih\,\delta t}\,V$, where $h=h_0+h_1$ is the single-particle Hamiltonian. On the open
chain of Fig.~\ref{fig:ricemele}, the Wannier centers are the eigenvalues of the projected
position operator $V^\dagger\hat x\,V$, with
$\hat x=\mathrm{diag}(0,1,\dots,L-1)$~\cite{marzari}. On the rings
of Figs.~\ref{fig:ricemelevar} and~\ref{fig:ricemelevar3N}, position is defined only modulo $L$,
and the Wannier centers $x_n$ follow from the eigenvalues $\lambda_n\propto e^{2\pi i x_n/L}$ of
the projected unitary~\cite{restaQuantumMechanicalPositionOperator1998}
\begin{equation}
\mathcal U=V^\dagger e^{2\pi i\hat x/L}\,V ;
\label{eq:Uproj}
\end{equation}
they are followed continuously in time.

The excited fraction in Fig.~\ref{fig:ricemelevar3N}(c) counts the electrons outside the
instantaneous lower band. With $P=VV^\dagger$ and $P_\mathrm{ad}$ the projector onto the $N$
lowest eigenvectors of $h_0(\theta)$, an instantaneous eigenmode $|u\rangle$ has occupation
$\langle u|P|u\rangle$, so the upper band holds $\mathrm{Tr}[(1-P_\mathrm{ad})P]$ electrons and
\begin{equation}
n_\mathrm{exc}=1-\frac{\mathrm{Tr}(P_\mathrm{ad}P)}{N}.
\label{eq:nexc}
\end{equation}
It vanishes on the adiabatic path, $P=P_\mathrm{ad}$, and, unlike the many-body fidelity,
measures the error per electron. Doubling or quadrupling the number of time steps changes
$n_\mathrm{exc}$ after five cycles by less than $10^{-3}$.

\subsection{Interacting chain (Figs.~\ref{fig:int_2vs14} and~\ref{fig:int_sizes})}
\label{app:num_int}

With $U$ and $V$ the state is no longer a Slater determinant, and Wannier centers of individual
electrons are not defined. We evolve the full many-body wave function in the occupation-number
basis, applying each step with the algorithm of Ref.~\cite{al-mohyComputingActionMatrix2011};
the benchmark with the exact CD term, a dense operator of rank two, is integrated with a
fourth-order Runge-Kutta scheme.

What remains defined is the position of all electrons together, through the Resta expectation
value $Z=\langle\psi|e^{2\pi i\hat X/L}|\psi\rangle$ of Eq.~\eqref{eq:resta}. Its phase is
accumulated step by step, $Q_{n+1}=Q_n+\arg(Z_{n+1}/Z_n)/2\pi$, which removes the $2\pi$
ambiguity. For a Slater determinant $Z=\det\mathcal U$, with $\mathcal U$ of
Eq.~\eqref{eq:Uproj}, so that
\begin{equation}
Q(t)=\frac{1}{L}\sum_n\big[x_n(t)-x_n(0)\big]:
\label{eq:QWannier}
\end{equation}
the pumped charge is the summed displacement of the Wannier centers of
Figs.~\ref{fig:ricemelevar} and~\ref{fig:ricemelevar3N}, divided by the length of the ring. An
ideal cycle moves each of the $L$ electrons by two sites, so $Q$ grows by 2 per cycle. $Q$ is
meaningful only while $|Z|$, which measures how well localized the center of mass of the
electrons is~\cite{restaElectronLocalizationInsulating1999}, is not small. With the
$\Delta E$-resolved CD term $|Z|\geq0.24$ for $L=8$, whereas without CD it drops below $0.01$;
the values of $Q$ without CD in Table~\ref{tab:int_results} therefore only signal that the pump
has failed.

In Fig.~\ref{fig:int_2vs14} the CD coefficients are computed at every time step; in
Fig.~\ref{fig:int_sizes} they are interpolated from the 40-point grid of
Sec.~\ref{sec:int_exact}. For $L=8$ the two procedures give end-of-cycle fidelities that agree
to $10^{-3}$, and doubling the number of time steps (quadrupling it at $T=2/t$) changes the
fidelities and $Q$ by at most $10^{-3}$.

\section{Variational CD for the Interacting Rice-Mele Model}
\label{app:interacting}

\subsection{Infinite-temperature action with interactions}
\label{app:S_inf}

We evaluate $S_\infty=\mathrm{Tr}\,G^2/\mathrm{Tr}\,\openone$ for the NN
ansatz~\eqref{eq:A_NN_spin}. The normalized trace is an average over all
occupation-number configurations of the chain with equal weight, i.e., every
spin-orbital is occupied independently with probability $1/2$.

\textit{Noninteracting part.} With the operators of Eq.~\eqref{eq:variational_ansatz}, the generalized force of
Sec.~\ref{subsec:variational} has components $g_x=p+q\cos k$, $g_y=q\sin k$, $g_z=r+w\cos k$, with
\begin{align}
p&=\dot t_1+2\Delta\alpha_1, & q&=\dot t_2-2\Delta\alpha_2,\nonumber\\
r&=\dot\Delta-2\alpha_1t_1+2\alpha_2t_2, & w&=2(\alpha_2t_1-\alpha_1t_2).
\label{eq:pqrw}
\end{align}
Transforming back to real space, $\cos k$ multiplying $\sigma_z$ becomes a hop to the same
sublattice of the neighboring cell, so
\begin{align}
G_0={}&\sum_{j,\sigma}\Big[p\,(c^\dagger_{j,A,\sigma}c_{j,B,\sigma}+\mathrm{h.c.})\nonumber\\
 &+q\,(c^\dagger_{j,B,\sigma}c_{j+1,A,\sigma}+\mathrm{h.c.})\nonumber\\
 &+r\,(n_{j,A,\sigma}-n_{j,B,\sigma})\nonumber\\
 &+\tfrac{w}{2}\,(c^\dagger_{j,A,\sigma}c_{j+1,A,\sigma}-c^\dagger_{j,B,\sigma}c_{j+1,B,\sigma}+\mathrm{h.c.})\Big].
\end{align}
In the configuration basis $G$ is a real symmetric matrix, so
$\mathrm{Tr}\,G^2=\sum_{c,c'}G_{cc'}^2$: every matrix element appears squared once. A hop
term $a(c^\dagger_{i\sigma}c_{j\sigma}+\mathrm{h.c.})$ has a matrix element $\pm a$ in the
half of the configurations in which exactly one of the two spin-orbitals is occupied, and
contributes $a^2/2$ to the average; a density term $r(n_{i\sigma}-n_{j\sigma})$ contributes
$r^2/2$; distinct terms never share a matrix element, or average to zero. Summing over the
two spins,
\begin{equation}
\frac{\langle G_0^2\rangle_\infty}{N}=p^2+q^2+r^2+\tfrac12w^2=\tfrac12S_{\mathrm{free}},
\end{equation}
where $S_{\mathrm{free}}=(2\pi)^{-1}\int dk\,2|\mathbf g|^2$ is the action of Sec.~\ref{subsec:variational}.

\textit{Interaction part.} By Eq.~\eqref{eq:dE},
$G_{\mathrm{int}}=\sum_{\text{bonds},\sigma}\alpha_{b}\,\Delta E\,(X+X^\dagger)$ has
matrix elements $\alpha_b\Delta E$ at exactly the positions of the corresponding hops in
$G_0$, where $\Delta E$ is evaluated in the occupations of the four sites around the bond
(the ``spectators'' of the hop). Write $\Delta E=(U-V)x+Vy$ with
$x=n_{a\bar\sigma}-n_{b\bar\sigma}$ and $y=n_{a'}-n_{b'}$. Over the four equally likely
states of the two opposite-spin orbitals, $x$ takes the values $0,+1,-1,0$, so
$\langle x\rangle=0$ and $\langle x^2\rangle=1/2$. Each outer site holds $n=0,1,1,2$
electrons with equal probability, so $\langle n\rangle=1$, $\langle n^2\rangle=3/2$, and
$\langle y\rangle=0$, $\langle y^2\rangle=2(\tfrac32-1)=1$. Since $x$ and $y$ involve
different orbitals, $\langle xy\rangle=\langle x\rangle\langle y\rangle=0$. Hence
\begin{equation}
\langle\Delta E\rangle=0,\qquad \langle\Delta E^2\rangle=\tfrac12(U-V)^2+V^2 ,
\end{equation}
for both bond types. The first result makes the cross term vanish,
$\langle G_0G_{\mathrm{int}}\rangle_\infty=0$, because the hop amplitude of $G_0$ is the
same in every spectator configuration while that of $G_{\mathrm{int}}$ averages to zero.
The second gives $\tfrac12\alpha_b^2\langle\Delta E^2\rangle$ per hop; with four hops per
unit cell (two bond types, two spins), $\langle G_{\mathrm{int}}^2\rangle_\infty/N=
(\alpha_1^2+\alpha_2^2)\langle\Delta E^2\rangle$. Together these give
Eq.~\eqref{eq:S_inf}. Using $\partial S_{\mathrm{free}}/\partial\alpha=8(M\alpha-b)$ from
Sec.~\ref{subsec:variational}, the condition $\partial S_\infty/\partial\alpha=0$ reads
$4(M\alpha-b)+2\langle\Delta E^2\rangle\alpha=0$, which is Eq.~\eqref{eq:Mc}.

\subsection{Free fermions}
\label{app:free_limit}

For $U=V=0$, $H=\sum_{k,\sigma}\psi^\dagger_{k\sigma}h(k)\psi_{k\sigma}$ with
$h(k)=\mathbf d\cdot\boldsymbol\sigma$. Let $\mathcal A(k)$ be the two-level gauge
potential of Eq.~\eqref{CDterm} and $\mathcal A=\sum_{k\sigma}\psi^\dagger_{k\sigma}\mathcal
A(k)\psi_{k\sigma}$. The commutator of two quadratic forms is the quadratic form of the
commutator, so $G=\sum_{k\sigma}\psi^\dagger_{k\sigma}\big(\dot
h+i[\mathcal A(k),h(k)]\big)\psi_{k\sigma}$. For the exact two-level gauge potential the
matrix in brackets is proportional to $h(k)$, so $[H,G]=0$, which by Appendix~\ref{app:variational_derivation} is the
defining condition of the exact many-body gauge potential. Equation~\eqref{eq:AGP_exact}
therefore reduces to Eq.~\eqref{CDterm}, applied to each $k$ and spin separately.

\subsection{Ground-state action}
\label{app:S_gs}

For $n\neq0$, $H|0\rangle=E_0|0\rangle$ and $\langle n|H=E_n\langle n|$ give
$\langle n|[\mathcal A,H]|0\rangle=(E_0-E_n)\langle n|\mathcal A|0\rangle$, hence
\begin{equation}
\langle n|G|0\rangle=\langle n|\dot H|0\rangle+i(E_0-E_n)\langle n|\mathcal A|0\rangle .
\end{equation}
The exact CD term $\dot\theta\mathcal A_\theta$ makes this vanish, so
$\langle n|\dot H|0\rangle=-i(E_0-E_n)\langle n|\eta\rangle$ with
$|\eta\rangle=\dot\theta\mathcal A_\theta|0\rangle$, and substituting back gives
Eq.~\eqref{eq:Gn0}. Dividing $|\langle n|G|0\rangle|^2$ by
$(E_n-E_0)^2$ and summing over $n\neq0$ gives Eq.~\eqref{eq:S_gs}. With
$\mathcal A=\sum_\mu\alpha_\mu O_\mu$, real $\alpha_\mu$, and the vectors
$|v_\mu\rangle=\mathcal Q O_\mu|0\rangle$ (note $\mathcal Q|\eta\rangle=|\eta\rangle$),
\begin{equation}
S_{\mathrm{gs}}=\sum_{\mu\nu}\alpha_\mu\alpha_\nu\,\mathrm{Re}\langle v_\mu|v_\nu\rangle
 -2\sum_\mu\alpha_\mu\,\mathrm{Re}\langle v_\mu|\eta\rangle+\langle\eta|\eta\rangle ,
\end{equation}
whose minimum is Eq.~\eqref{eq:Mgs}. The target $|\eta\rangle$, Eq.~\eqref{eq:eta}, is
obtained by solving $(H-E_0)\,x=\mathcal Q\,\dot H|0\rangle$ for $x$ (we use the
MINRES algorithm) and setting $|\eta\rangle=-ix$; then the $14\times14$ system~\eqref{eq:Mgs}
is solved for the coefficients.

\subsection{Interaction-energy classes}
\label{app:classes}

The projectors in Eq.~\eqref{eq:A_dE} are products of occupation projectors. For the hop
of a spin-$\sigma$ electron from $b$ to $a$, the value of $x$ is selected by
\begin{align}
P^{(x=+1)}&=n_{a\bar\sigma}(1-n_{b\bar\sigma}),\nonumber\\
P^{(x=-1)}&=(1-n_{a\bar\sigma})n_{b\bar\sigma},\nonumber\\
P^{(x=0)}&=1-P^{(x=+1)}-P^{(x=-1)},
\end{align}
and the value of $y$ by products of the site projectors
$[n_s=0]=(1-n_{s\uparrow})(1-n_{s\downarrow})$,
$[n_s=1]=n_{s\uparrow}+n_{s\downarrow}-2n_{s\uparrow}n_{s\downarrow}$ and
$[n_s=2]=n_{s\uparrow}n_{s\downarrow}$ on the outer sites $s=a',b'$. Then
$P_\varepsilon=\sum_{(x,y):\,(U-V)x+Vy=\varepsilon}P^{(x)}P^{(y)}$.
Table~\ref{tab:dE_classes} lists the resulting energies for $U=8t$, $V=4t$.

\begin{table}[h]
\caption{Interaction energy $\varepsilon=(U-V)x+Vy$ (in units of $t$) of a NN hop for
$U=8t$, $V=4t$, for all values of $x=n_{a\bar\sigma}-n_{b\bar\sigma}$ and
$y=n_{a'}-n_{b'}$. Because $U-V=V$, the 15 combinations take only seven distinct values.}
\label{tab:dE_classes}
\begin{ruledtabular}
\begin{tabular}{c|ccccc}
$x\,\backslash\,y$ & $-2$ & $-1$ & $0$ & $+1$ & $+2$\\
\hline
$-1$ & $-12$ & $-8$ & $-4$ & $0$ & $+4$\\
$0$  & $-8$  & $-4$ & $0$  & $+4$ & $+8$\\
$+1$ & $-4$  & $0$  & $+4$ & $+8$ & $+12$\\
\end{tabular}
\end{ruledtabular}
\end{table}

\subsection{Fitted coefficients}
\label{app:coeffs}

\begin{figure}[!htb]
\centering
\includegraphics[width=\columnwidth]{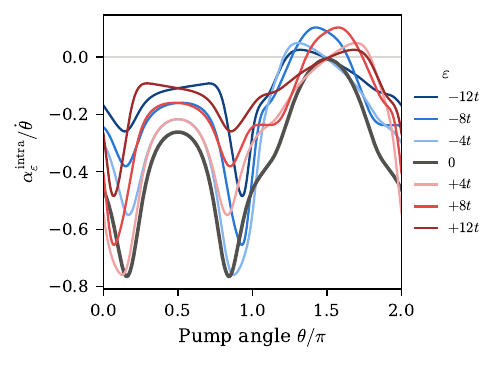}
\caption{Fitted intracell coefficients $\alpha^{\mathrm{intra}}_\varepsilon/\dot\theta$
of the $\Delta E$-resolved ansatz, Eq.~\eqref{eq:A_dE}, over one cycle for $L=12$, with
one curve per interaction-energy class $\varepsilon$. The intercell coefficients follow
from $\alpha^{\mathrm{inter}}_\varepsilon(\theta)=\alpha^{\mathrm{intra}}_\varepsilon(\theta+\pi)$.}
\label{fig:int_coeffs}
\end{figure}

Figure~\ref{fig:int_coeffs} shows the intracell coefficients of the $\Delta E$-resolved
ansatz obtained from $S_{\mathrm{gs}}$ for $L=12$. The coefficients vary smoothly along the cycle, so a grid of 40 values of $\theta$ with
periodic spline interpolation suffices. They decrease with $|\varepsilon|$: the largest
value of $|\alpha^{\mathrm{intra}}_\varepsilon|/\dot\theta$ along the cycle is $0.77$ for
$\varepsilon=0$, $0.76$ for $|\varepsilon|=4t$, $0.66$ for $|\varepsilon|=8t$ and $0.49$
for $|\varepsilon|=12t$. This dependence is weaker than the ratio of the interaction
energies alone would suggest, and the $\varepsilon=0$ coefficient is not small, because
the energy denominators of Eq.~\eqref{eq:AGP_exact} also contain the kinetic and
staggered-potential energies and the ground state is a superposition of many
configurations.

The coefficients for $L=8$ differ from those for $L=12$ by $0.05$ in root mean square over
all classes and grid points. The largest differences, up to $0.3$, occur in the classes
$|\varepsilon|\geq8t$. Their operators carry only about 12\% of the total weight
$\sum_\mu\|\mathcal Q O_\mu|0\rangle\|^2$ on the ground state ($L=8$), so they are less
tightly constrained by the fit.

\bibliographystyle{apsrev4-2}
\bibliography{references}

\end{document}